\documentstyle[preprint,aps]{revtex}
\begin{document}
\tightenlines
\newcommand{\het}{$^{\rm 3}$He}
\newcommand{\hef}{$^{\rm 4}$He}
\newcommand{\nf}{$^{\rm 14}$N}
\newcommand{\of}{$^{\rm 15}$O}
\newcommand{\bes}{$^{\rm 7}$Be}
\newcommand{\ba}{$^{\rm 8}$B}
\newcommand{\nalpha}{$\alpha$}
\newcommand{\nbeta}{$\beta$}
\newcommand{\ngamma}{$\gamma$}
\newcommand{\betap}{$\beta^+$}
\newcommand{\nue}{$\nu_{\rm e}$}
\newcommand{\num}{$\nu_{\rm m}$}
\newcommand{\nut}{$\nu_{\rm t}$}
\newcommand{\hetp}{\het$^+$}
\newcommand{\hethet}{\het(\het,2p)\hef}
\newcommand{\hetd}{\het(d,p)\hef}
\newcommand{\hethef}{\het(\nalpha, \ngamma)\bes}
\newcommand{\besp}{\bes(p,\ngamma)\ba}
\newcommand{\nde}{$^{13}$N(e$^+$\nue)$^{13}$C}
\newcommand{\ofe}{$^{15}$O(e$^+$\nue)$^{15}$N}
\newcommand{\ecm}{E$_{\rm cm}$}
\newcommand{\elab}{E$_{\rm lab}$}
\newcommand{\pmev}{MeV$^{\rm -1}$}
\newcommand{\pqcm}{cm$^{\rm -2}$}
\newcommand{\qcm}{cm$^{\rm 2}$}
\newcommand{\pccm}{cm$^{\rm -3}$}
\newcommand{\ps}{s$^{\rm -1}$}
\newcommand{\E}[1]{10$^{\rm #1}$}
\newcommand{\mum}{$\mu$m}
\newcommand{\mus}{$\mu$s}
\title{The Cross Section of \hethet\  measured at Solar Energies}
\author{
THE LUNA COLLABORATION \\
M.~Junker$^{1,2}$,
A.~D'Alessandro$^1$,
S.~Zavatarelli$^3$,
C.~Arpesella$^1$,
E.~Bellotti$^4$,
C.~Broggini$^5$,
P.~Corvisiero$^3$ \footnote{corresponding author: P. Corvisiero,
Dipartimento di Fisica and INFN, Via Dodecaneso 33, 16146 Genova, Italy,
Tel.: +39 10 3536299, Fax.: +39 10 313358, E--Mail: Corvi@ge.infn.it},
G.~Fiorentini$^6$,
A.~Fubini$^7$,
G.~Gervino$^8$,
U.~Greife$^2$,
C.~Gustavino$^1$,
J.~Lambert$^9$,
P.~Prati$^3$,
W.S.~Rodney$^9$,
C.~Rolfs$^2$,
F.~Strieder$^2$,
H.P.~Trautvetter$^2$,
D.~Zahnow$^2$}
\address{
$^1 $) Laboratori Nazionali Gran Sasso, Assergi\\
$^2 $) Institut f\"ur Experimentalphysik III, Ruhr--Universit\"at Bochum\\
$^3 $) Universit\'{a} di Genova, Dipartimento di Fisica and INFN, Genova\\
$^4 $) Universit\'{a} di Milano,  Dipartimento di Fisica and INFN, Milano\\
$^5 $) INFN, Padova                  \\
$^6 $) Universit\'{a} di Ferrara, Dipartimento di Fisica and INFN, Ferrara\\
$^7 $) ENEA, Frascati and INFN, Torino\\
$^8 $) Politecnico di Torino, Dipartimento di Fisica and INFN, Torino\\
$^9 $) Georgetown University, Washington}
\maketitle
\begin{abstract}
We report  on the results of the \hethet\ experiment
at the underground accelerator facility LUNA (Gran Sasso). For the
first time the lowest
projectile energies utilized for the cross section measurement correspond  to
energies below the center of the solar Gamow peak ($E_{\rm 0}$=22 keV). 
The data provide no evidence for the existence of a hypothetical resonance in
the energy range investigated. 
Although no extrapolation is needed anymore (except for energies at the
low-energy tail of the Gamow peak), the data must be corrected for the effects
of electron screening, clearly observed the first time for the \hethet\
reaction.
The effects are however larger than expected and not understood, leading
presently to the largest uncertainty on the quoted $S_{\rm b}(E_{\rm 0})$
value for bare nuclides (=5.40 MeV b).
\end{abstract}
\pacs{26., 24., 26.65.+t, Nucl. Astrophys., Nucl. React., Solar
Neutrinos} 
\pagebreak
\section{Introduction}
\label{secintro}
Accurate knowledge of thermonuclear reaction rates is important 
$\cite{fow84,rol88}$
in understanding the generation of energy, the luminosity  of  neutrinos,
and the synthesis of elements in stars. Due to the Coulomb barrier
(height $E_{\rm c}$) of
the entrance channel, the reaction cross  section  $\sigma(E)$ drops nearly
exponentially with decreasing energy  $E$.  Thus  it  becomes increasingly
difficult to measure $\sigma(E)$ and to deduce the  astrophysical $S(E)$
factor defined by the equation $\cite{rol88}$

\begin{equation}
\sigma(E)= \frac{S(E)}{E}\,{exp(-2\,\pi\, \eta)}, \label{yielddef1}
\end{equation}
with the Sommerfeld parameter given by
$2\, \pi\, \eta=31.29\, Z_1\, Z_2(\mu/E)^{1/2}$.
The quantities $Z_1$ and $Z_2$  are  the 
nuclear charges of the interacting particles  in  the  entrance channel,
$\mu$ is the reduced mass (in units of amu),  and  $E$  is  the  center-of-mass
energy  (in   units  of  keV).  Although   experimental  techniques  have
improved $\cite{rol88}$ significantly over the  years  to extend
$\sigma(E)$ measurements
to lower energies, it has not  yet been  possible  to  measure  $\sigma(E)$
within the thermal energy region in stars. This region is  determined by
the Gamow energy window $E_0 \pm \delta E_0$
(the Gamow peak) for a given  stellar
temperature and lies far below the height of the Coulomb
barrier, approximately at $E_{\rm 0}/E_{\rm c} = 0.01$.
Instead,
the observed $\sigma(E)$ data at  higher  energies had to  be
extrapolated to thermal energies. As always in physics, such an extrapolation
into the unknown can lead to considerable uncertainties. 

The low-energy  studies of thermonuclear reactions in a laboratory at
the  earth's  surface are hampered predominantly by the effects of cosmic
rays in the detectors. Passive shielding around the detectors provides
a reduction of gammas and neutrons from the environment, but  it  produces
at the same time an increase of gammas and neutrons due to the cosmic-ray
interactions  in  the  shield itself. A 4$\pi$ active shielding can
only  partially  reduce  the  problem  of cosmic-ray background. An excellent
solution  is  to  install  an  accelerator  facility in a laboratory deep
underground \cite{fio95}. As a pilot project, a 50 kV accelerator facility 
has been
installed $\cite{arp91,gre94,gre95}$ in  the  Laboratori Nazionali 
del Gran Sasso (LNGS), where
the flux of cosmic-ray muons is reduced by a factor \E{6} \cite{mac90}.
The LUNA pilot
project  (LUNA = Laboratory for Underground  Nuclear Astrophysics)
was designed  primarily  for a renewed study of the \hethet\
reaction ($Q$ = 12.86 MeV) in the  energy  range of the
solar Gamow peak ($E_0 \pm \delta E_0$ = $21.9 \pm 6.2$ keV) for a central 
star temperature of $T = 15.5\times10^6$ K. The reaction
is a member of the  hydrogen  burning  proton--proton (pp) chain $\cite{rol88}$,
which is
predominantly  responsible   for  the  energy  generation   and  neutrino
luminosity $\cite{bah92}$ of the sun.
So far, the reaction has been studied down to about 25 keV (section II)
but there remains the possibility of a narrow resonance at lower energies.

The hypothesis of a low energy resonance was first advanced  
$\cite{fow72,fet72}$ as a
solution to the solar neutrino puzzle, which in those days was regarded as
a deficit of $^{8}$B neutrinos. For this purpose a resonance with
$E_{\rm R}{\simeq}$ 21 keV and ${\Gamma}<$ 6 keV was considered 
$\cite{fet75}$ as the most favourable case.
Experimental data available at that time were not inconsistent with the
existence of a resonance with $E_{\rm R}$=15-20 keV and ${\Gamma}<$ 2 keV 
$\cite{cas93}$.
As more data on solar neutrinos became available, it became clear that
the deficit of $^{7}$Be neutrinos is stronger than that of $^{8}$B neutrinos.
It was shown $\cite{cas93}$ that such a pattern of suppression occurs if
$E_{\rm R}{\simeq}$ 21 keV.

Such a resonance level in $^6$Be has
been sought $\cite{rol88}$ without success by various indirect routes, and it
is also not
predicted by most nuclear-structure theories. However, the  existence  of
this hypothetical resonance can be positively  dismissed  only  by direct
measurements at the required low energies (i.e. within the solar Gamow peak).

For nuclear reactions studied  in the laboratory,
the  target  nuclei and the projectiles are usually in the form of neutral
atoms/molecules and ions, respectively. The  electron clouds surrounding the
interacting nuclides act as a screening potential: the projectile effectively
sees a reduced Coulomb barrier.  This  in turn leads to a higher cross section,
$\sigma_{\rm s}(E)$, than would  be  the case for bare nuclei, 
$\sigma_{\rm b}(E)$,  with an exponential  enhancement  factor 
$\cite{ass87,bra90}$

\begin{equation}
f_{\rm lab}(E) = \sigma_{\rm s}(E)/
\sigma_{\rm b}(E) \simeq \exp(\pi\eta\, U_{\rm e}/E), \label{yielddef2}
\end{equation}  
where $U_{\rm e}$ 
is the electron-screening potential energy 
(e.g. $U_{\rm e} \simeq  Z_1\cdot Z_2 \cdot e^2/R_{\rm a}$
approximately, with $R_{\rm a}$ an atomic radius). 
It should be pointed out that for a stellar plasma the value of
$\sigma_{\rm b}(E)$ must  be known because the screening in  the  plasma
can  be quite
different from that in laboratory studies $\cite{ric95}$, 
and $\sigma_{\rm b}(E)$
must be explicitly included in each situation. Thus, a good understanding of
electron-screening effects is needed to arrive at reliable  $\sigma_{\rm b}(E)$
data  at low energies.
Low-energy studies of several fusion reactions involving light  nuclides showed
$\cite{gre95,pra94,lan96}$
indeed  the  exponential enhancement of the cross section at low energies.
The observed enhancement (i.e. the value of $U_{\rm e}$)
was -- in all cases -- close to or higher than the  adiabatic  limit
derived from atomic-physics models. An exception are the previous 
\het+\het\ data (section II),
which show apparently a flat $S(E)$ curve down  to $E=25$ keV, 
although the
effects of electron screening should have enhanced the data at 25 keV by about
a factor 1.2 for the adiabatic limit $U_{\rm e}$ = 240 eV. Thus, improved
low-energy data are particularly desirable for this reaction.

We report here on such new data obtained by the LUNA Collaboration within
the solar Gamow peak.
Preliminary results, which have been published $\cite{arp96}$, 
are superseded by the present report.
\section{The \hethet\ reaction}
The \hethet\ reaction represents in the exit channel a 3-body-breakup: if the 
breakup is direct, one should observe a continuous energy distribution  of
the ejectiles described by phase-space considerations; if the breakup follows
a sequential process, the energies of the ejectiles are described by 2-body
kinematics. Experiments have shown $\cite{bac65,dwa71,kra87}$ 
that at energies below $E = 1$ MeV
the reaction proceeds predominantly via a direct mechanism and that the
angular distributions approach isotropy with decreasing energy. The $S(E)$ 
energy dependence observed by various groups 
$\cite{bac65,dwa71,kra87,dwa74,wan66}$ represents a consistent
picture (Fig.\ \ref{fig1}). 
The only exception is the earliest experiment $\cite{goo51}$ where
$S(E)$ is lower by a factor 3 to 5 compared to the other experiments; the
discrepancy is most likely caused by target problems ( \het\ trapped in an Al
foil).

The absolute $S(E)$ values of refs. $\cite{dwa71}$ and $\cite{kra87}$ 
(as well as those from the 
present work, section VI) agree - at overlapping energy regimes - within 
experimental uncertainties, while those of refs. $\cite{dwa74}$ 
and $\cite{wan66}$ are lower by about
25\%, suggesting a renormalization of their absolute scales. However, in 
view of the relatively few data points and their relatively large uncertainties
-- in comparison to the other data sets -- it has been suggested $\cite{ade97}$
to omit these data, without significant loss of information.
We verified that no changes in the $S(E)$ fit are appreciable (within 1\%)
by including or excluding the data sets from $\cite{dwa74}$ and $\cite{wan66}$.

A reaction mechanism was suggested $\cite{may68}$ at low energies, 
in which a neutron
tunnels from one \het\ to the other, unimpeded by the Coulomb barrier, up to
a radial distance where the nuclei overlap appreciably. In this model, a 
diproton remains and subsequently fissions into 2 protons. The calculated
energy dependence of the $S(E)$ factor described well the data (dotted curve 
in Fig.\ \ref{fig1}), thus providing confidence in the extrapolation using a 
polynomial function (solid curve in Fig.\ \ref{fig1}):

\begin{equation}
S(E)  = S(0) + S'(0)E + 0.5\,S''(0)E^2   
      = 5.3 - 3.7 \, E + 1.9 \, E^2 {\rm(MeV\ b)}  
\label{yielddef3}
\end{equation}
\section{The LUNA facility}
Technical details of the LUNA setup have been reported $\cite{gre94}$. 
Briefly, the
50 kV accelerator facility (Fig.\ \ref{fig2}) 
consisted of a duoplasmatron ion source,
an extraction/acceleration system, a double-focusing 90$^{\rm o}$ analysing
magnet (with adjustable pole faces), a windowless gas-target system, and a beam
calorimeter.

The energy spread of the ion source was less than 20 eV, the plasma potential
energy deviated by less than 10 eV from the voltage applied to the anode, and
the emittance of the source was 2 cm rad eV$^{1/2}$. The ion source provided
a stable beam current of about 1 mA over periods of up to 4 weeks. 

The high voltage (HV) of the accelerator was provided by a power supply,
which has a typical ripple of $5 \times 10^{-5}$, a longterm stability of 
better
than $1\times10^{-4}$, and a temperature coefficient of better than 
$1.5\times 10^{-4}$ /K.
The air-conditioned laboratory was kept at a temperature of $T = 21^{\rm o}$ C
and a relative humidity of $H=30$\%. The HV of the accelerator was 
measured with
a resistor chain, contained in an air-tight plexiglas tube, and a digital
multimeter. The resistor chain was built as a voltage divider, with fifty 20
M$\Omega$ resistors and one 100 k$\Omega$ resistor (temperature coefficient = 
$1\times 10^{-5}$/K). The multimeter (with a long term stability of 
$5\times 10^{-5}$ per
year) provided the numerical value of the HV measured across the 100 k$\Omega$
resistor. This HV-measuring device was calibrated at the PTB in Braunschweig
(Germany) at $T = (20 \pm 1)^{\rm o}$C and $H = (35 \pm 10)$\%  to a 
precision of $5\times 10^{-5}$. 

The beam entered the target chamber of the differentially pumped gas-target
system (3 pumping stages) through apertures of high gas flow impedance
(C to A in Fig.\ \ref{fig2}) 
and was stopped in the beam calorimeter. The gas pressure
in the target chamber, $p_{\rm o}$, was measured with a Baratron capacitance
manometer to an accuracy of better than 1\%. This measurement was absolute
and independent of the type of gas used. For \het\ gas (99.9\% enriched in
\het\ ) of $p_{\rm o}$ = 0.50 mbar pressure, the system reduced the pressure
to $1\times 10^{-3}$, $1\times 10^{-4}$, and $1\times 10^{-5}$ 
mbar in the three  upstream 
pumping stages; a similar pressure reduction was observed for other 
$p_{\rm o}$ values. The gas composition in the target chamber was monitored
with a mass spectrometer. The \het\ gas was compressed by Roots blowers,
cleaned efficiently using a zeolite adsorption trap (cooled to liquid
nitrogen temperature), and fed back into the target chamber (gas 
recirculation). The pressure $p_{\rm o}$ was kept at a constant value using
a needle valve in combination with an electronic regulation unit. As noted
above, the main pressure drop occurred across the entrance aperture A
(7 mm diameter, 40 mm length, 230 mm distance from aperture B). It was shown
that the pressure in the target chamber was essentially unmodified by the
gas flow through the entrance aperture A; thus, the geometrically extended
target zone was characterized by a nearly static pressure. Beam-heating
effects on the gas density are expected to be less than 0.5\% $\cite{goe85}$
for a maximum $p_{\rm o} = 
0.50$ mbar \het\ target pressure and the 500 $\mu$A maximum \het$^+$
beam current.

The beam current in the target area was determined to an accuracy of 3\%
using the beam calorimeter (with a constant temperature gradient). The
calorimeter was placed (Fig.\ \ref{fig3}) at such a distance $d$
from the center of aperture A
($d=32.2 \pm 0.1$ cm) that angle straggling of the incident beam in the gas
resulted in a beam profile smaller than the 200 mm$^{2}$ active area of the
calorimeter.

The LUNA facility was equipped with an interlock system, which allowed 
the system to run without an operator on site. The duty time of the facility
in the chosen running conditions was about 90\%, with a weekly service
time of 8 h. As the typical beam current in the target area was about 400
$\mu$A, a weekly charge of about 200 Cb could be accumulated on the 
target. 
\section{The detection setup}
The detection setup for the \hethet\ studies had to fulfill the following 
requirements:
\begin{enumerate}
\item a high absolute efficiency, in view of the expected reaction rates
(yields) of about 1 event/day and less;
\item a high rejection of natural radioactivity in the detectors, in the
target chamber facing the detectors, and from the surrounding rocks at LNGS
(mainly gammas); tests at LNGS have shown that commercial Si detectors 
exhibited an intrinsic radioactivity level, which was about 200 times higher 
than the above reaction yield;
\item a high rejection of electronic noise, in view of the needed running
times of several weeks per energy point;
\item a clear separation of the reaction products from those of \hetd\ 
($Q$ = 18.35 MeV), due to  deuterium contamination in the \het\
beam (as HD$^{+}$ molecules of mass 3) and in the gas target (found to be
smaller).
This contaminant reaction has a cross
section one millionfold higher than that of \hethet\ at $E_{\rm lab}= 40$ keV, 
mainly due to the barrier ratio $E_{\rm c}$ (d+ \het )/$E_{\rm c}$ (\het\ +
\het ) = 0.56, and thus  extremely small deuterium contaminations (of order 
10$^{-7}$) can lead to sizeable event rates.
\end{enumerate}
In order to optimize the detection setup and to understand the resulting 
spectra for quantitative analyses, a Monte Carlo program $\cite{arp95}$
was written
to simulate the experiment under realistic conditions. The Monte Carlo
program produces energy and time spectra of the ejectiles as well as
absolute yields, which could be compared directly with data. Various 
quantitative tests of the Monte Carlo predictions have been carried out
successfully $\cite{gre94,gre95,arp95}$. 
A Monte Carlo simulation $\cite{arp95}$ of the proton 
spectrum from both reactions
at a beam energy of 50 keV is shown in Fig.\ \ref{fig4}: 
a Ni foil of \het\  and 
\hef\ stopping thickness (20 $\mu$m) was placed in front of a 1000 $\mu$m
thick Si detector (maximum commercially available thickness at the
time of the experimental project: the desirable
thickness was 1400 $\mu$m to stop the 14.7 MeV protons from \hetd\ ); a 
deuterium contamination of 10$^{-6}$ was assumed. One sees the continuous 
energy distribution of the protons from \hethet\ and the peak(s) plus
low-energy tail of protons from \hetd\ (due to incomplete  stopping in the
detector). Both proton spectra overlap appreciably leading to large 
uncertainties in the yield analysis of the \hethet\ events; thus, the single
spectra of Si detectors did not represent a viable solution. In addition, 
singles spectra would not reject the background due to natural radioactivity
and electronic noise. By the requirement of proton-proton coincidences
between detectors -- placed on opposite sides of the beam axis -- a unique
signature of the \hethet\ events could be obtained $\cite{kra87}$
in the coincidence
spectra; also the background events (radioactivity and electronic noise)
could essentially be eliminated. However, the price for this unique 
signature is a reduction of the absolute efficiency, by at least one order 
of magnitude compared to analysis via single spectra.
Therefore, this approach was not followed; instead the present approach
utilized ${\Delta}E$-E
detectors, in which single proton signals were detected in coincidence between
the ${\Delta}E$ and E detectors of the telescope.
In the detection setup (Fig.\ \ref{fig3}), designed according to the indications
of Monte Carlo simulations, four ${\Delta}$E-E telescopes
(placed in a rectangular target chamber) were arranged around the beam axis: 
two opposite telescopes each at a distance of 2.7 cm from the beam axis
and the two other telescopes each at 3.7 cm; the distance from the center
of the entrance aperture A to the center in front of the telescopes was
$(8.3 \pm 0.1)$ cm. 
The different distance from the center of the up-down and left-right
couples of telescopes was due to mechanical constraints. 
Each telescope consisted of transmission surface barrier
silicon detectors with a 0.25 \mum\ thick Al layer deposited on both sides
of the detectors. The $\Delta $E and E detectors both had an active square
area of 2500 mm$^{2}$; the $\Delta $E (and E) detector had a thickness
of 140 \mum\ (and 1000 \mum ) and an energy resolution of 105 keV
(and 55 keV) at $E_{ \alpha }$ = 5.5 MeV. A mylar foil (1.2 \mum\ thick)
and an Al foil (1.5 \mum\ thick) were placed in front of each telescope;
they stopped the intense elastic scattering yield and shielded the detectors
from beam induced light.
This double shielding was proved, in various background runs, to be effective
and free from pinhole effects.
The detectors were maintained permanently at low temperature
(about -20$^{\rm o}$C) using a liquid recirculating cooling system.
The \hef\ ejectiles from \hethet\ ($E_{ \alpha }$  = 0 to 4.3 MeV) and \hetd\
($E_{ \alpha }$  = 3.7 MeV) were stopped in the $\Delta $E detectors, while
the ejected protons from \hethet\ ($E_{\rm p}$ = 0 to 10.7 MeV) and \hetd\
($E_{\rm p}$ = 14.9 MeV) left signals in both the $\Delta $E and E 
detectors of a given telescope (coincidence requirement). Fig.\ \ref{fig5} 
shows the
identification matrix of one $\Delta $E-E telescope simulated with the 
Monte Carlo program at $E_{\rm lab}$= 50 keV and assuming a deuterium 
contamination of 10$^{ -6}$: the matrix reveals a clear separation of the
events from both reactions.

Standard NIM electronics were 
used in connection with the four telescopes. The signals were
handled   and   stored   using  a   CAMAC  multiparametric   system,  which
allowed for  on-line  as  well  as   for  play-back  data  analyses.
Signals from two or more detectors were considered coincident within a time
window of 1 $\mu$s, making negligible the rate of random coincidences.
A pulser was permanently used in all detectors to check for dead time and
electronic stability. The
acquisition system  also stored concurrent  information on experimental
parameters (such as  ion  beam  current/charge,  accelerator  high  
voltage,  and
gas  pressure  in  the   target   chamber)  via  CAMAC  scalers.  Controls
have also  been  implemented to stop  data  acquisition  if  the  beam  was
lost  or  to  reject  an  event  if  an  error in  the  data  transmission
occured.
The system  handles safely a data flux of about 500 events/s.

In the analysis of the data, the accepted events from \hethet\ had to
fulfill the following 3 conditions:
\begin{enumerate}
\item the events had to arise from proton-induced $\Delta$E-E coincidences
in a given telescope; the coincidence requirement of each telescope
essentially eliminated events due to 
natural  radioactivity of the detectors themselves and of surrounding 
materials;
\item the events had to lie in a closed region of the
$\Delta$E-E plane (Fig.\ \ref{fig6}), whose borders were
first deduced by Monte Carlo simulations and then fixed in  order  to  cut
the electronic  noise;
\item  a proton event
had to occur in only one given E-detector; events which
triggered more than one E-detector were rejected;
this condition reduced significantly events due to the residual
electronic noise, but it also rejected some p-p coincidences.

\end{enumerate}

The chosen $\Delta$E-E energy region together with the anticoincidence
requirement led to an absolute detection efficiency of $(7.55 \pm 0.15)$\%
as determined by the Monte Carlo program. In the $\Delta$E-E region of
events from \hethet\ (Fig.\ \ref{fig6}), no background events were ever observed
at LNGS since the installation of the equipment (January 1994) during
several ``no beam" and/or ``no target" background measurements (up to two 
months running) as well as during a 10 day run with a \hef\ beam and     
a \hef\ target gas. At the Bochum laboratory (earth surface) a background
rate of about 10 events/day was found with the same setup. With the 
discussed operating conditions the setup sensitivity in terms of cross 
section values is better than 10$^{-14}$ b.
Although  the  selected  regions in the spectra of the
telescopes allow for a clear separation  of the events from both reactions
(Fig.\ \ref{fig6}), a few protons from the contaminant reaction \hetd\
can hit the detectors
near the edges of their active volumes  loosing  only  a fraction of their
energy and  thus leading  to  a  background rate in the
$\Delta$E-E region of the
\hethet\  events. The probability of such events was investigated by
the  Monte  Carlo  program  as  well  as  by  direct  measurements   using
projectiles with $Z/A$ = 0.5 (selected by the 90$^{\rm o}$ analysing magnet). The
ratio between the background events in the \het\ -\het\ region 
and those in the
clearly separated d -\het\ region (= monitor) turned out to be
(0.40 $\pm$ 0.04)\%. Thus a deuterium contamination equal to 10$^{-7}$ allows
$\sigma$(E) measurements as low as $E$ = 15 keV, with a signal-to-background
ratio equal to 4.
During  the  reported experiments  the deuterium contamination 
d/\het\ ranged between
$5\times10^{-8}$ and $5\times10^{-6}$.
\section{Effective beam energy and absolute cross section}
For the isotropic emission of the protons in \hethet\  (section II),
the number of counts $dN(z)$ per unit of time arising from a differential
length $dz$ of the extended \het\ gas target is given by the expression

\begin{equation}
dN(z)=N_{\rm t} \cdot N_{\rm b} \cdot \sigma(E(z))\cdot  
\eta(z)\cdot dz, \label{yielddef4}
\end{equation}
where $N_{\rm t}$ is the \het\ target density in units of atoms/cm$^{3}$
(derived from the measured target pressure $p_{\rm o}$, with a precision
better than  1\%\ ), $N_{\rm b}$  is the number of \het\ projectiles per
unit of time (derived from the beam calorimeter, with a precision of
3\% ), and  $\eta(z)$ is the absolute detection efficiency of all four
telescopes including geometry and the acceptance criteria discussed in 
section IV. The efficiency also takes into account 
that two protons are
produced per reaction. Introducing the stopping power $\epsilon$ (i.e.
the energy loss per unit length), equation (4)  can be rewritten in the
form

\begin{equation}
 dN(E) = N_{\rm t} \cdot N_{\rm b} \cdot \sigma(E) \cdot
 \eta(E) \cdot  \epsilon (E)^{-1} \cdot dE.
 \label{yielddef5}
\end{equation}
 
The total number of counts for the full target length $L=(32.2 \pm 0.1)$ cm 
(from the
center of the aperture A to the beam calorimeter) is then given by

\begin{equation}
 N = N_{\rm t} \cdot N_{\rm b} \cdot \int _{L} \sigma(E) \cdot
 \eta(E) \cdot  \epsilon (E)^{\rm -1} \cdot dE.
 \label{yielddef6}
\end{equation} 

For the case of a thin target, i.e. introducing an effective beam energy
$E_{\rm eff}$ corresponding to the mean value of the projectile energy
distribution in the detection setup (see below), one arrives at

\begin{equation}
 N = N_{\rm t} \cdot N_{\rm b} \cdot \sigma(E_{\rm eff}) \cdot
  \int _{L} \eta(E) \cdot  \epsilon (E)^{\rm -1} \cdot dE,
 \label{yielddef7}
\end{equation}   
where the values for $E_{\rm eff}$ and the integral were derived from
the Monte Carlo program.

At subcoulomb energies a precise knowledge of the effective beam energy
associated with observed yields is as important as the yield 
measurements themselves. For this reason, all Monte Carlo predictions 
have been thoroughly
tested $\cite{arp95}$. Here they are compared with a simple
``hand" calculation.
Let us consider the case of an incident \het\
energy of $E_{\rm lab}$ = 50.00 keV. The absolute energy is known to a 
precision of $\pm$ 2.5 eV, its long term stability is $\pm$ 5.0 eV, and the
Gaussian energy spread (FWHM) is 20 eV (section III). When the beam
passes through the gas target system filled with \het\ at the standard
pressure of $p_{\rm o}$ = 0.30 mbar, the beam loses energy in the 3
pumping stages and in the target chamber. 
For the calculation of this
energy loss, we used stopping-power values given by the TRIM
program $\cite{zie95}$. It should be pointed out that experimental energy-loss
data  $\cite{gol92}$ at the relevant low energies were found to be consistent
with those from the TRIM program, within the experimental error of
10\%. We adopted this error in our analyses. With $\epsilon =
(7.0 \pm 0.7)10^{-15}$ eV/atom/cm$^{2}$ at $E_{\rm lab} = 50$ keV
one finds an energy loss of 
$\Delta E_{1}=(2.2 \pm 0.2)$ eV 
in the 3 pumping stages and $\Delta E_{2}$ = (430 $\pm$ 43) eV
over the (8.3 $\pm$ 0.1) cm distance from the center of 
aperture A to the center
of the target chamber (detector location): $\Delta E_{\rm tot}$ =  (432
$\pm$ 43) eV. The quoted error arises predominantly from the uncertainty
in $\epsilon$ . This leads to an effective energy at the center of the
target chamber of $E_{\rm lab} = (49.568 \pm 0.043)$ keV.
The telescopes see an effective target length of $(7.0 \pm 0.1)$ cm
(see Fig.\ \ref{fig8}). The energy loss over this target length (i.e. the target
thickness) is $\Delta E_{3} = 365$ eV with an estimated error of $\pm$ 36 eV.
For a  constant $S(E)$ factor
the cross section drops by 11.2 \% over this target thickness.
Assuming a linear decrease in cross section 
over this region $\cite{rol88}$, the effective
energy evaluation gives
$E_{\rm lab} = (49.572 \pm 0.045)$ keV, or a center-of-mass
energy $E_{\rm eff} = (24.786 \pm 0.023)$ keV. The error on
$E_{\rm eff}$ transforms into a $\pm$ 1.5\% uncertainty on 
$\sigma(E_{\rm eff}$).
These estimates are consistent with Monte Carlo calculations simulating
precisely the detection setup (Fig.\ \ref{fig3}) and the acceptance criteria of
\het-\het\ events. The effective energy,
as deduced from the mean energy of the distribution
shown in Fig.\ \ref{fig7}, 
correspond to $E_{\rm lab**}= (49.568 \pm 0.043)$ keV.
The spread of the energy distribution is 
determined by various effects: 
a) the energy straggling $\cite{rol88,bes81}$ at the center of the target, 
 $\delta_{E_{1}} = 125$ eV; b) the
thermal Doppler width $\cite{rol88}$ at room temperature, 
$\delta_{E_{2}}= 51$ eV;
c) the energy spread over the target length seen by the detectors, 
$\delta_{ \Delta E_{3}}= 105$ eV. The total energy
spread $\delta_{E}$ is therefore
171 eV, with an estimated error of $\pm 20$ eV.
This is in good agreement with the 189 eV spread predicted by the Monte
Carlo program (Fig.\ \ref{fig7}).
This procedure was applied for all beam energies and gas pressures.
It should be noted that the 10\% systematic uncertainty in the projectiles
energy loss in the target induces only  a 0.1 -- 0.2 \% error on $E_{\rm eff}$
corresponding to a 1.0 -- 3.5 \% uncertainty in the $S(E)$ values.

>From the geometry of the four square telescopes and the target length 
$L= 32.2$ cm one expects an absolute efficiency over this length of the
order of 10\%; the Monte Carlo simulations yield $(7.55 \pm 0.15)$\%. The
geometrical setup efficiency was tested using an $\alpha$-source, which
was moved along the beam axis.
This test was possible, due to mechanical constraints, only recently 
with a new setup (to be used in future work)
where eight 1000 \mum\ thick detectors were placed in
two consecutive boxes 
(each containing 4 detectors around the beam axis)
all at the same distance from the beam axis.
The results (Fig.\ \ref{fig8}) are in good agreement
with the corresponding Monte Carlo simulations  (solid
curves in Fig.\ \ref{fig8}).

To test the reliability of the corrections applied in the data analyses,
we performed measurements at $E_{lab} = 50.00$ keV using different gas
pressures: $p_{\rm o}$ = 0.15, 0.30 and 0.50 mbar; for each pressure
the target thickness, the energy loss, and the detection efficiency are
different. The resulting $S(E)$ values (Table I) are fully compatible 
with each other, well inside the experimental uncertainties.

During another test period the detection setup was changed: the detector box
was moved 5.0 cm closer to the beam calorimeter extending the total
target length from 32.2 cm to 37.2 cm. The target pressure was fixed 
at $p_{\rm o}$ = 0.50 mbar. In this setup a significant decrease (of the
order of 25\%) in reaction yield was predicted by the Monte Carlo
simulations compared to the standard setup and standard pressure
($p_{\rm o}$ = 0.30 mbar), due to the lower effective energy.
The $S(E)$ results (Table 1) are -- within the
experimental uncertainties -- in good agreement with values deduced for the
standard setup.

\section{Results}
Table I and Fig.\ \ref{fig9} summarize the \hethet\ results obtained
until December 1996
with the LUNA setup at the 50 kV underground accelerator facility (LNGS).
The lowest counting rate was 3 events per day at $E=$ 20.76 keV. The preliminary
data  $\cite{arp95}$ have been completely re-analysed and have been
integrated with the results of other new measurements.
The data obtained at higher energies  (450 kV accelerator in Bochum)  with
the LUNA setup $\cite{gre94}$
are also included for completeness. 
Previous literature data obtained
at $E= 24.51$ to 1080 keV $\cite{dwa71,kra87}$ are also
shown in Fig.\ \ref{fig9}. 
The LUNA data have been obtained
at energies within the solar Gamow peak, i.e. below the 22 keV
center of this peak, and represent the first measurement of an important
fusion cross section at energies near the center of the Gamow peak. No evidence
of the  hypothetical resonance can be found in the covered energy range.
The observed energy dependence $S(E)$, for bare nuclides $S_{\rm b}(E)$
and shielded nuclides $S_{\rm s}(E)$ 
was parametrized using the expressions $\cite{rol88,ass87}$
\begin{eqnarray}
S_{\rm b}(E) & = & S_{\rm b}(0) + S_{\rm b}'(0)E + 0.5\, S_{\rm b}''(0)E^2  \\
S_{\rm s}(E) & = & S_{\rm b}(E) \exp(\pi \eta U_{\rm e}/E),
\end{eqnarray}
where $S_{\rm b}(0)$, $S_{\rm b}'(0)$, $S_{\rm b}''(0)$, and $U_{\rm e}$ 
are fit parameters.
The data set shown in Fig.\ \ref{fig9} was fitted, in the energy region
between 20.7 and 1080 keV, using three methods:
\begin{itemize}
\item[a)] first $S_{\rm b}(0)$, $S_{\rm b}'(0)$, $S_{\rm b}''(0)$ were 
obtained considering data for
$E \geq 100$ keV only; then $U_{\rm e}$ was deduced from all the data 
and the fixed $S_{\rm b}(E)$ expression;
\item[b)] all the four parameters were allowed to change and the 
whole data set was
fitted; the resulting curves for $S_{\rm b}(E)$ and $S_{\rm s}(E)$
are shown in Fig.\ \ref{fig9} as dashed and solid curves, respectively;
\item[c)] $U_{\rm e}$ was fixed at 240 eV (i.e. the value 
calculated in the adiabatic limit) while $S_{\rm b}(0)$, 
$S_{\rm b}'(0)$ and $S_{\rm b}''(0)$  were allowed to vary.
\end{itemize}

The results are summarized in Table II. The three methods give compatible
$S_{\rm b}(E)$ values while $U_{\rm e}$ changes significantly: the
methods a) and b) give $U_{\rm e}$ values higher than the adiabatic
limit (240 eV), consistent with observations in other fusion reactions.
It should be noted that fitting the LUNA data only with method b), a
screening potential of 490$ \pm$ 30 eV is obtained.
The difference between observed and predicted $U_{\rm e}$ values is
not understood at present. 
Therefore we adopted $S_{\rm b}(E_{\rm 0})$ from the theoretical adiabatic
screening (method c) and derived a further error corresponding to the extreme
cases $U_{\rm e}=0$ eV ($S_{\rm b}(E_{\rm 0})=5.7$ MeV b) 
and $U_{\rm e}=432$ eV ($S_{\rm b}(E_{\rm 0})=5.1$ MeV b):
\[
S_{\rm b}(E_{\rm 0})= 5.40 \pm 0.05  \pm 0.30
 \pm 0.30 ) {\rm (MeV b)} 
\]
where the first two errors arise 
from statistical and systematical uncertainties, respectively, and the last 
error from the lack of understanding of electron screening.

\section{A hypothetical resonance at lower energies}
The difference between the LUNA data in the 20.7-24.8 keV region and
$S(E)$ calculated in the adiabatic limit (i.e. fixing
$U_{\rm e}$=240 eV), can be used to estimate upper limits for the strength of
a hypothetical resonance located at lower energies. 
For each measured point we calculated an "excess" S-factor,
$S_{\rm exc}(E)$ from:

\begin{equation}
S_{\rm exc}(E) =S_{\rm meas}(E)-S_{\rm ad}(E),
\label{sexc}
\end{equation}      

where $S_{\rm meas}$(E) are the data given in Table I corrected for a 240 eV screening
potential and $S_{\rm ad}$ comes from the fit c) in Table II discussed above. The excess
cross section, integrated in the measured energy interval (20.7 - 24.8 keV),
gives the integral yield:

\begin{equation}
Y_{\rm exc} =  (2.0 \pm 0.5   ) 10^{-15 } MeV \cdot b.
\label{excvalue}
\end{equation}      

We may compare this value 
with the tail of a low-energy resonance integrated over the same energy range.
For this purpose, we adopted the following procedure: the resonant cross 
section, $\sigma_ {\rm res}$(E), was described  according to the usual 
Breit-Wigner expression:

\begin{equation}
\sigma_ {\rm res}(E) = \pi\ \lambdabar^{2} \omega \Gamma\
 _{\rm a}(E) \Gamma\  _{\rm b}(E) \frac{1} {(E - E_ {\rm R})^{2} + 
(\Gamma\ /2)^{2}},
\label{sigres}
\end{equation} 

where $\lambdabar $ is the De Broglie wavelength, $\omega$  is 
the statistical factor, $E_ {\rm R}$ is
the resonance energy  and $\Gamma = \Gamma_{\rm a} +
 \Gamma_{\rm b}$ is the resonance total width.
Due to the high energies of the protons in the exit channel, the partial
width of the resonance decay, $\Gamma_{\rm b}(E)$,
was assumed to be energy-independent.
For the partial width of the entrance channel, $\Gamma_{\rm a}(E)$, we used 
the expression $\cite{rol88}$: 

\begin{equation}
\Gamma_{\rm a,l}(E) = (\frac {2E} {\mu})^{0.5} \frac {2 \hbar } {R _{\rm n}}
P_{\rm l}(E,R_{\rm n}) \theta_{\rm l}^{2},
\label{gammadia}
\end{equation} 

where $\mu$ = 1.507 a.m.u. is the reduced mass,$R_{\rm n}$ = 4 fm is the
 nuclear radius and 
the dimensionless reduced width $\theta_{\rm l}^{2}$ 
of the nuclear state has an upper limit of unity ($\theta_{\rm l}^{2}$ 
$\leq$ 1, Wigner limit). The penetrability $P_ {\rm l}(E,R _{\rm n})$ is
given by the equation:

\begin{equation} 
P_ {\rm l}(E,R _{\rm n}) = \frac {1} {F_{\rm l} ^{2}(E,R_{\rm n}) +
G_{\rm l} ^{2}(E,R_{\rm n})},
\label{pdielle}
\end{equation}  

and was calculated for s-waves (l=0) using the approximations for the Coulomb
wave functions, $F_ {\rm l}$ and $G_ {\rm l}$, given in $\cite{abr65}$.
It turns out that $\Gamma_{\rm a} \ll$ 1 keV at E $\leq$ 25 keV 
( for $\theta_{\rm l}^{2}$ = 1) and thus $\Gamma  \simeq \Gamma_{\rm b}$.

Equations (12),(13) and (14) lead to the expression:

\begin{equation}
\sigma_ {\rm res}(E) = A \frac {1} {E^{0.5}} 
\frac {1} {F_ {\rm l}^{2}(E,R_{\rm n}) +
G_ {\rm l}^{2}(E,R_{\rm n})}
\frac {\Gamma} {(E - E_ {\rm R})^{2} + (\Gamma\ /2)^{2}},
\label{sigfat}
\end{equation}   

where the constant A contains the parameters $\mu$, $R_{\rm n}$ and
$\theta_{\rm l}^{2}$. Its value  can
 be determined by the comparison with the experimental value
$Y_{\rm exc}$:

\begin{equation}
\int_{\rm 20.7} ^{\rm 24.8} \sigma_ {\rm res}(E) dE = Y_{\rm exc}.
\label{integr}
\end{equation}  

With this functional relation $\theta_{\rm l}^{2}$ is known for fixed values
of the resonance energy $E_{\rm R}$ and of the total width $\Gamma$.
It is then  possible to evaluate the reaction rate 
$<\sigma v>_{\rm res}$
of this hypothetical resonance and 
compare it
with the non-resonant rate $<\sigma v>_{\rm SSM}$ used in Standard Solar Model
(SSM) calculations.
Fig.\ \ref{fig10} shows the calculated ratio
$r =<\sigma v>_{\rm res}/<\sigma v>_{\rm SSM}$ 
as a function of $E_{\rm R}$ for some values of the total resonance width
$\Gamma$, assuming a central temperature of the
sun of 15 $10^{6}$ K.
The dotted curve in Fig.\ \ref{fig10} shows the reaction rate ratio in the
Wigner limit ($\theta_{\rm l}^{2}$ = 1), which was calculated using 
equations (15) and (16).

The $<\sigma v>_{\rm res}$ values obtained with this procedure are
upper limits for the reaction rate due to the hyphotetical 
resonance. At energies $E_{\rm R} \leq$ 9 keV one concludes
that the presence  of a resonance 
cannot account for even a partial nuclear solution of the solar
neutrino puzzle $\cite{cas93,ber95}$.                                           
The same conclusion applies in the energy region between 9 and 20 keV for 
resonance 
widths $\geq$ 2 keV. Room seems still available for a very narrow resonance
in this interval where only direct measurements could rigorously
dismiss (or confirm) its existence.
\section{Conclusions}

For the first time  the \hethet\ fusion reaction has been studied in an
energy region deep inside the Gamow peak, i.e. down to 20.7 keV.
This goal has been achieved at the new facility for nuclear astrophysics
(LUNA) built in the underground laboratory of Gran Sasso (Italy).The results
obtained so far show that the $S(E)$ energy dependence is consistent with the
predictions based on an extrapolation from higher energies. The presence of a
low energy resonance in the \hethet\ reaction, which could have strong effects
on the ``solar neutrino problem", does not emerge from the new data.
In the near future the LUNA collaboration will extend the measurements down
to $E = 17$ keV: the foreseen running time here is one year. Definite
conclusions with respect to the expected solar
neutrino fluxes have to await the results of these experiments. For
this next phase a new detection set-up, designed to reduce the background
induced by deuterium contamination in the beam and in the gas target, has
been developed and is now in a testing phase.

The electron screening effect, which produces an exponential
enhancement of $S(E)$ at 
low energies, can be observed in the new \hethet\ data. This is a second
important result since previous data did not
show a clear evidence of the enhancement (Fig.\ \ref{fig1}). 
The screening potential
seems to be higher than the value predicted by the adiabatic model of the
interaction between projectile and atomic electron clouds. It should be
noted that the adiabatic approximation, among other electron screening
models, is that giving the highest screening potential. A better
quantitative determination of the screening potential will be possible
when the LUNA collaboration will reach the lowest energy scheduled for the
\hethet\ experiment. Hopefully ,a complete and clear picture of the screening
effect will be  possible when data collected at very low energies and for
other fusion reactions are available. 
This is one of the future goals of the LUNA collaboration.

The excess of the screening potential could also be explained as
a tail of a narrow resonance lying in the not measured low-energy region.
Upper limits for the strength of such a resonance have been calculated
assuming the adiabatic limit for the screening potential.
Unfortunately, these calculations cannot exclude the existence of a 
narrow resonance (width < 2 keV) at energies between 9 and 20 keV
with a strength high enough to give a sizeable contribution to
the \hethet\  reaction rate in our sun.

\section{Acknowledgements}
We are indebted with the Director and all the staff of the Laboratori
Nazionali del Gran Sasso for the hospitality and the support offered to
this experiment
\clearpage
\begin{figure}
\caption{ Astrophysical $S(E)$ factor of the \hethet\ reaction as obtained 
in previous work $\protect\cite{dwa71,kra87}$ (section II). 
The solid curve is a polynomial fit to
the data and the dotted curve a theoretical calculation 
$\protect\cite{may68}$ normalized to $S(0)$ = 5.1 MeV b\label{fig1}.}
\end{figure} 
\begin{figure}
\caption{Schematic diagram  of the 50 kV LUNA accelerator facility (see
also Figs. 2 and 3 of \protect\cite{gre94}).\label{fig2} }
\end{figure}
\begin{figure}
\caption{Schematic diagram of the rectangular target chamber including the
detection setup (telescopes) and the beam calorimeter; the given lengths
and diameters ($\phi$) are in units of mm (see also Figs.$\ \protect\ref{fig4}$
and $\ \protect\ref{fig5}$ 
in \protect\cite{gre94}).\label{fig3} }
\end{figure}
\begin{figure}
\caption{Monte Carlo simulation of the protons from \hethet\ and \hetd\
detected in a 1000 \mum\ thick Si detector (covered with a 20 \mum\ thick
Al foil) at $E_{\rm lab}$ = 50 keV and in the setup shown in 
Fig. $\ \protect\ref{fig3}$. A
deuterium contamination in the \het\ beam of d/\het\ = 10$^{-6}$ was
assumed.\label{fig4}  }
\end{figure}
\begin{figure}
\caption{Monte Carlo simulation of the $\Delta$E-E identification matrix
for the protons from \hethet\ and \hetd\ at $E_{\rm lab}$ = 50 keV in 
the setup shown in Fig. $\ \protect\ref{fig3}$. 
A deuterium contamination in the \het\ beam
of d/\het\ = 10$^{-6}$ was assumed.\label{fig5} }
\end{figure}
\begin{figure}
\caption{$\Delta$E--E identification matrix of one telescope (obtained
in the setup of Fig. $\ \protect\ref{fig3}$) 
at $E_{\rm lab}$ = 50 keV and $p_{\rm o}$ = 0.30
mbar. The \het+\het\ and d+\het\ selected regions are shown; note the
beam-induced electronic noise at the left vertical edge of the matrix.
\label{fig6}}
\end{figure}
\begin{figure}
\caption{Absolute detection efficiency in a new modified setup 
(to be used in future work) with 
eight E detectors (arranged in two consecutive boxes)
as observed with an $\alpha$-source ($E_{ \alpha }$ = 5.5 MeV),
which was moved along the beam axis (distance $d_{\rm A}$ measured 
from the center
of the entrance aperture A). All the detectors have the same distance
(2.7 cm) from the beam axis. The first section (four detectors) of
the setup is nearly equivalent 
to the configuration in Fig.$\ \protect\ref{fig3}$. The target length 
seen by one box (7 cm) is the same as in the setup in Fig.
$\ \protect\ref{fig3}$. 
The solid curves show the results of 
Monte Carlo simulations.\label{fig8} }
\end{figure}
\begin{figure}
\caption{Monte Carlo simulation of the energy distribution of the \het\
projectiles (incident energy $E_{\rm lab}$ = 50.00 keV, \het\ gas
pressure $p_{\rm o}$ = 0.30 mbar) leading to detected/accepted events
in the setup of the 4 telescopes (Fig.$\ \protect\ref{fig3}$).
The mean energy of the
distribution is $E_{\rm lab**}$ = (49.568 $\pm$ 0.043)keV and the
energy spread is $\delta$E = 189 eV.\label{fig7}}
\end{figure}
\begin{figure}
\caption{The  $S(E)$ factor of \hethet\ from the previous work
(Fig.$\ \protect\ref{fig1}$) and the present work (Table 1). 
The dashed and solid curves 
represent $S_{\rm b}(E)$ and $S_{\rm s}(E)$, respectively. 
The solar Gamow peak is shown in
arbitrary units. The upper right corner inset zooms on the underground
LUNA data. \label{fig9}}
\end{figure}
\begin{figure}
\caption{The  ratio $<\sigma v >_{\rm res}/< \sigma v >_{\rm SSM}$ of the
resonance and non resonance reaction as a function of the resonance
energy $E_{\rm R}$ for some values of the total width  $\Gamma$.
The dashed curve shows the upper limit of this ratio considering the
Wigner limit ($\theta_{\rm l}^{2}$ = 1). \label{fig10}}
\end{figure}      
\newpage
\begin{table}[h]
\caption{S(E) factor of \hethet\ }
\begin{tabular}{rrrrrrr}
%
%
\multicolumn{1}{c}{Energy \tablenotemark[1]}  &
\multicolumn{1}{c}{Charge \tablenotemark[2]}  &
\multicolumn{1}{c}{Counts \tablenotemark[3]}  &
\multicolumn{1}{c}{Background \tablenotemark[4]} &
\multicolumn{1}{c}{S(E)} &
\multicolumn{1}{c}{$\Delta$S$_{\rm stat}$ \tablenotemark[8]} &
\multicolumn{1}{c}{$\Delta$S$_{\rm sys}$ \tablenotemark[9]} \\
\cline{5-7}
\multicolumn{1}{c}{  (keV) }    &
\multicolumn{1}{c}{   (Cb) }     &           &     &
\multicolumn{3}{c}{(MeV b)}\\
\hline
%
  91.70 \tablenotemark[6] 
& 0.0373 &   16479   & \tablenotemark[5]  &  5.15 & 0.11  &   0.21  \\
  86.51 \tablenotemark[6]
& 0.0301 &    8931   & \tablenotemark[5]  &  5.23 & 0.11  &   0.22  \\
  81.50 \tablenotemark[6]
& 0.0544 &    8378   & \tablenotemark[5]  &  5.33 & 0.12  &   0.22  \\
  76.29 \tablenotemark[6]
& 0.209  &   22879   & \tablenotemark[5]  &  5.32 & 0.11  &   0.22  \\
  71.22 \tablenotemark[6]
& 0.0945 &    5012   & \tablenotemark[5]  &  5.43 & 0.14  &   0.26  \\
  66.06 \tablenotemark[6]
& 0.0615 &    2304   & \tablenotemark[5]  &  5.43 & 0.15  &   0.26  \\
  61.06 \tablenotemark[6]
& 0.238  &    3562   & \tablenotemark[5]  &  5.41 & 0.14  &   0.26  \\
  55.94 \tablenotemark[6]
& 0.257  &    2251   & \tablenotemark[5]  &  5.50 & 0.16  &   0.29  \\
  50.64 \tablenotemark[6]
& 0.825  &    2870   & \tablenotemark[5]  &  5.63 & 0.14  &   0.31  \\
  45.82 \tablenotemark[6] 
& 0.784  &    1087   & \tablenotemark[5]  &  6.14 & 0.23  &   0.39  \\
  24.80 \tablenotemark[7] \tablenotemark[11]
&   204.2        & 128       &  14.5      &5.96 & 0.62 &0.23\\
  24.70 \tablenotemark[7] \tablenotemark[10]
&   344.0        & 424       &  47.5      &6.23 & 0.37 &0.24\\
  24.52 \tablenotemark[7] \tablenotemark[12]
&    49.5        & 100       &   7.7      &7.10 & 0.79 &0.31\\
  24.30 \tablenotemark[7] \tablenotemark[13]
&    68.5        & 119       &  10.9      &6.90 & 0.72 &0.37\\
  24.25 \tablenotemark[7] \tablenotemark[10]
&    99.8        &  93       &   5.0      &6.66 & 0.74 &0.26\\
  23.70 \tablenotemark[7] \tablenotemark[10]
&   140.4        &  96       &   4.2      &6.87 & 0.74 &0.27\\
  23.21 \tablenotemark[7] \tablenotemark[10]
&   122.9        &  73       &   8.2      &7.50 & 1.02 &0.30\\
  23.15 \tablenotemark[7] \tablenotemark[13]
&    32.7        &  28       &   2.9      &6.82 & 1.47 &0.42\\
  22.82 \tablenotemark[7] \tablenotemark[13]
&   139.3        & 103       &  12.5      &7.21 & 0.84 &0.39\\
  22.78 \tablenotemark[7] \tablenotemark[10]
&   307.0        & 101       &   6.2      &5.97 & 0.64 &0.24\\
  22.33 \tablenotemark[7] \tablenotemark[13]
&   113.9        &  57       &   3.7      &7.27 & 1.05 &0.40\\
  22.28 \tablenotemark[7] \tablenotemark[10]
&   233.5        &  59       &   7.3      &5.85 & 0.89 &0.24\\
  21.75 \tablenotemark[7] \tablenotemark[10]
&   373.9        &  77       &   2.4      &7.63 & 0.91 &0.31\\
  21.23 \tablenotemark[7] \tablenotemark[10]
&   416.4        &  60       &   6.5      &7.15 & 1.06 &0.29\\
  20.76 \tablenotemark[7] \tablenotemark[10]
&  1044.9        & 107       &  17.1      &6.80 & 0.82 &0.28\\
\end{tabular}
\tablenotetext[1]
{Effective center-of-mass energy derived from the absolute energy of the
ion beam and  Monte  Carlo
calculations  (including the  energy loss of the projectiles in the target
gas and the effects of the extended gas-target and detector geometries).}
\tablenotetext[2]
{Deduced from the beam calorimeter (3\% accuracy).}
\tablenotetext[3]
{Observed events in  the  \het -\het\  region (Fig.\ \ref{fig4}).}
\tablenotetext[4]
{Background events in the \het -\het\ region equal to  0.40\% of the observed
counts in the d-\het\ region of the spectra (Fig.\ \ref{fig4}), 
which must be subtracted from the counts in column 3.}
\tablenotetext[5]
{Negligible background.}
\tablenotetext[6]
{Data obtained with the LUNA setup at the 450 kV accelerator in  Bochum
\ \cite{gre94}, with the standard setup and target gas pressures ranging 
from 0.15 to 0.30 mbar.}
\tablenotetext[7]
{Data obtained with the LUNA setup  and accelerator facility at LNGS
(energy spread $=$ 20 eV); the chosen energy steps are of the order of the
energy loss of the beam in the extended gas target.}
\tablenotetext[8]
{Statistical error (one standard deviation) including  counting
statistics and apparative variations (pressure (1\%), beam power (1.3\%),
and temperature (1\%)).}
\tablenotetext[9]
{Systematical error (one standard deviation) including uncertainties in
pressure (1\%), beam power (3\%), efficiency (2\%), beam energy (200 eV
c.m. at Bochum, negligible at LNGS), and  energy loss (10\%).}
\tablenotetext[10]
{Target gas pressure equal to 0.30 mbar, standard detection setup position.}
\tablenotetext[11]
{Target gas pressure equal to 0.15 mbar, standard detection setup position.}
\tablenotetext[12]
{Target gas pressure equal to 0.50 mbar, standard detection setup position.}
\tablenotetext[13]
{Target gas pressure equal to 0.50 mbar, modified detection setup position.}
\end{table}
\newpage
\begin{table}
\caption{$S_{\rm b}(E)$ factor of \hethet\ for bare nuclides and electron
screening potential energy $U_{\rm e}$, from different fitting
procedures in the 20.7-1080 keV region. A description of the adopted
procedures is given in the text.}
\begin{tabular}{rrrrrr}
%
%
\multicolumn{1}{c}{$S_{\rm b}(0)$}  &
\multicolumn{1}{c}{$S_{\rm b}'(0)$}  &
\multicolumn{1}{c}{$S_{\rm b}''(0)$}  &
\multicolumn{1}{c}{$U_{\rm e}$} &
\multicolumn{1}{c}{$\chi^2$} &
\multicolumn{1}{c}{method} \\
\multicolumn{1}{c}{MeV b}  &
\multicolumn{1}{c}{b}  &
\multicolumn{1}{c}{b/MeV}  &
\multicolumn{1}{c}{eV} &
\multicolumn{1}{c}{} &
\multicolumn{1}{c}{} \\  
\hline
 $5.1 \pm 0.1$ & $-2.6 \pm 0.7$ & $2.0 \pm 1.3$ & $432 \pm 29$  & 0.93 & a\\%
 $5.30 \pm 0.08$ & $-3.6 \pm 0.6$ & $3.8 \pm 1.1$ & $323 \pm 51$  & 0.85 & b\\%
 $5.40 \pm 0.05$ & $-4.1 \pm 0.5$ & $4.6 \pm 1.0$ & 240     & 0.87 & c\\%
\end{tabular}
\end{table}
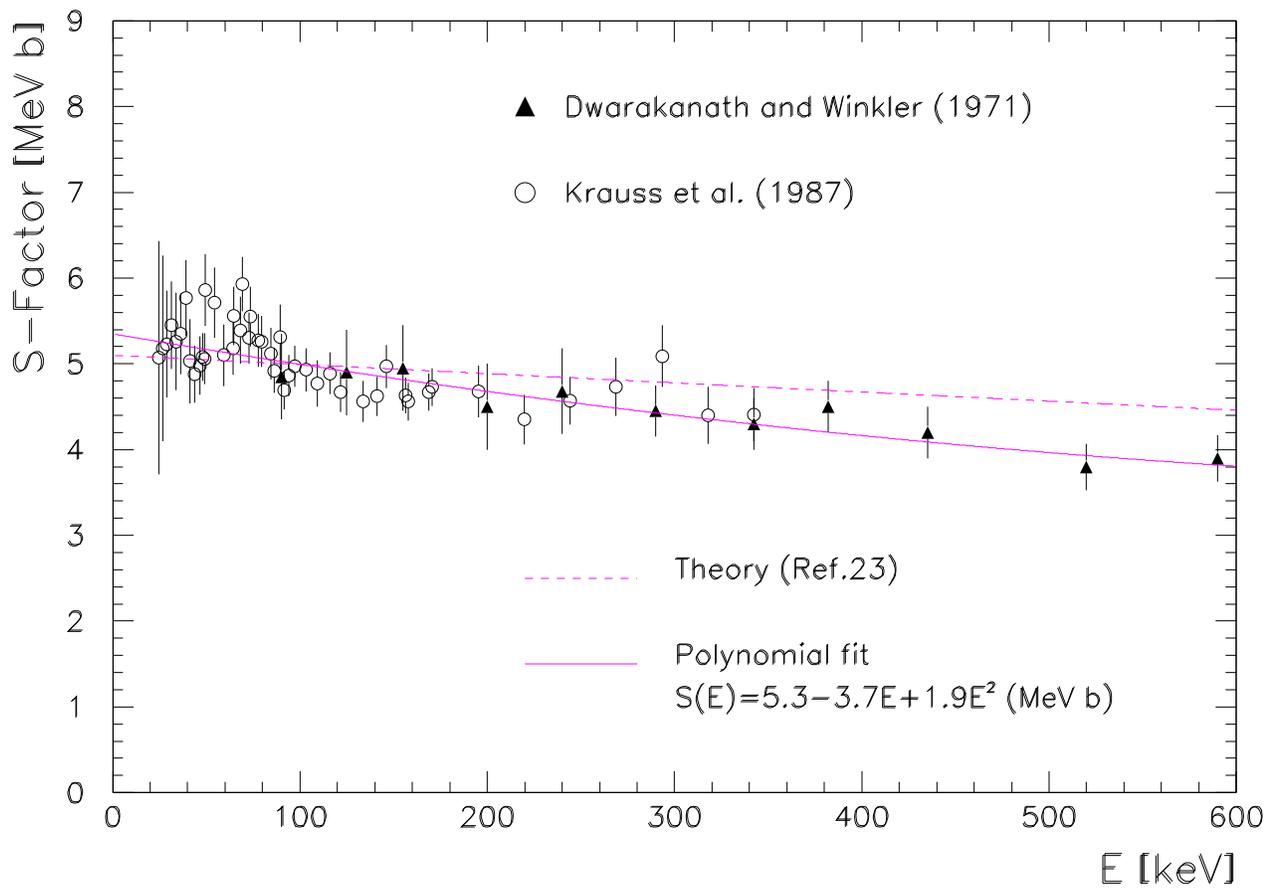
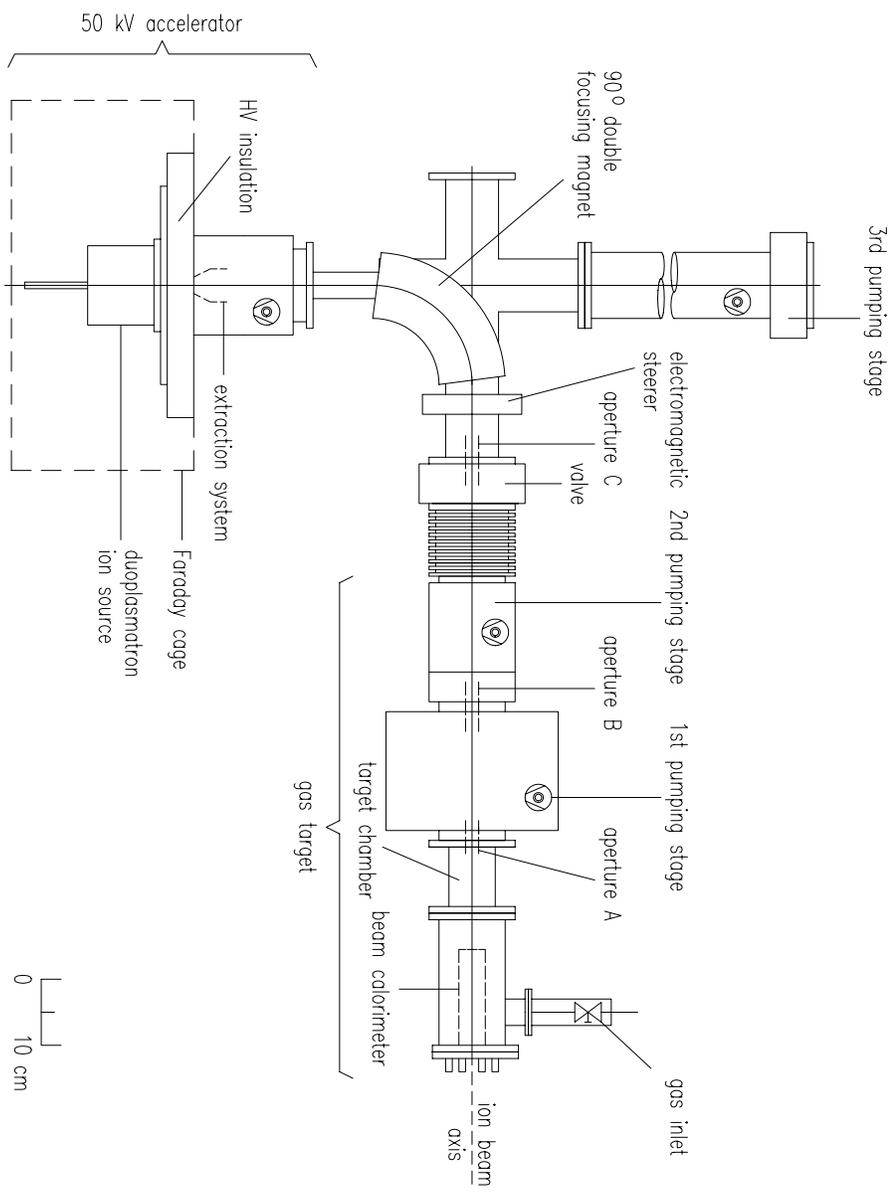
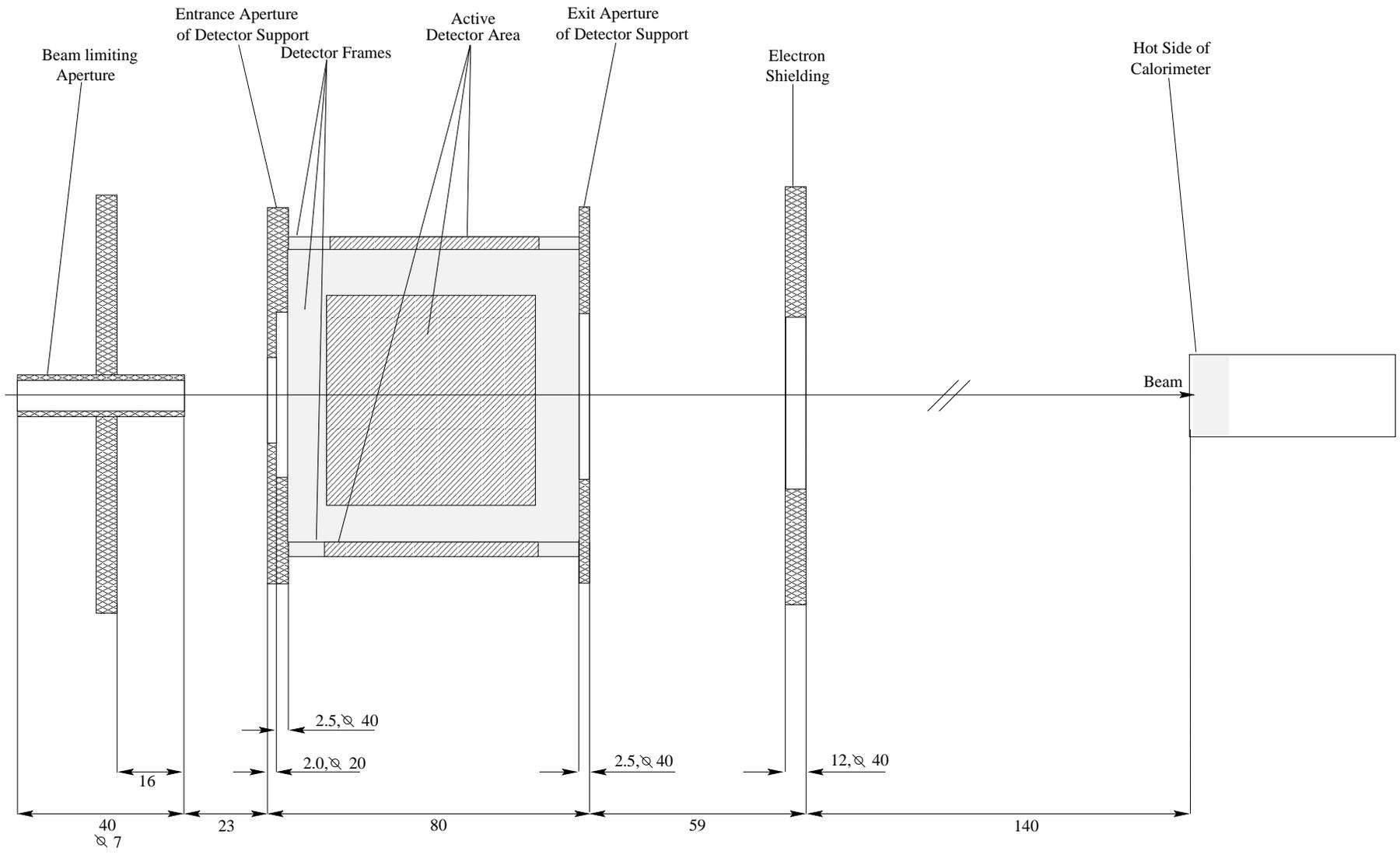
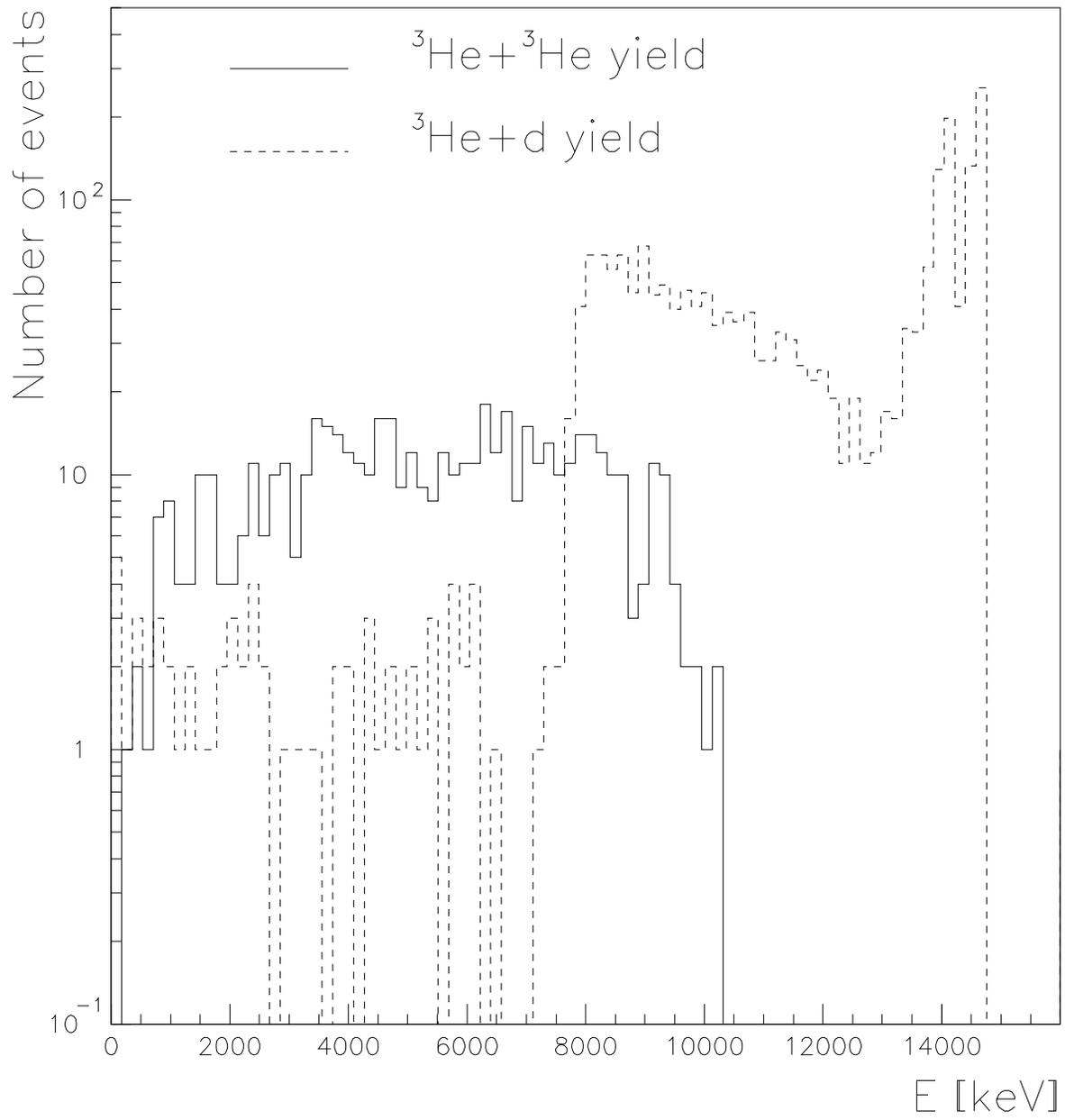
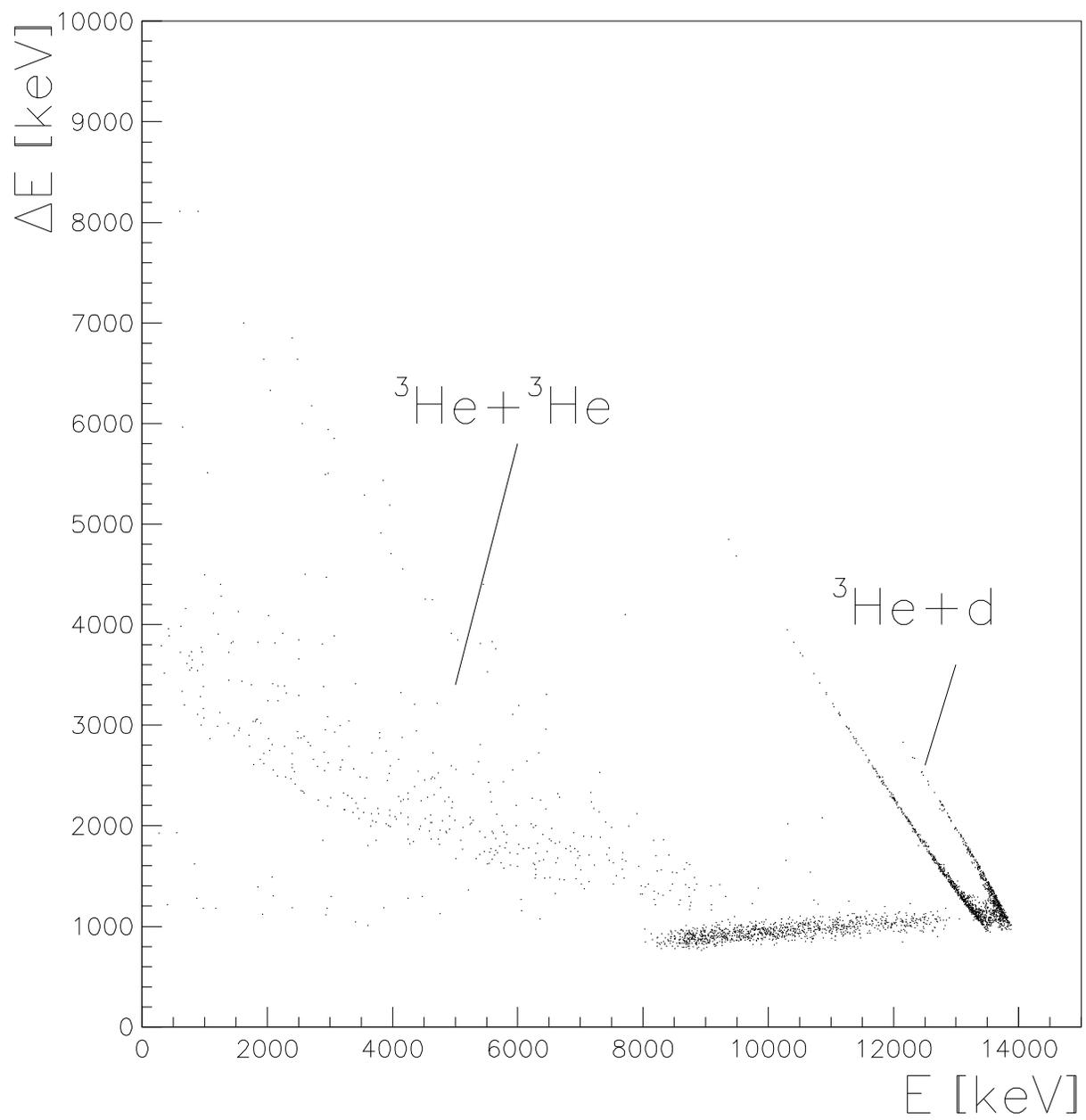
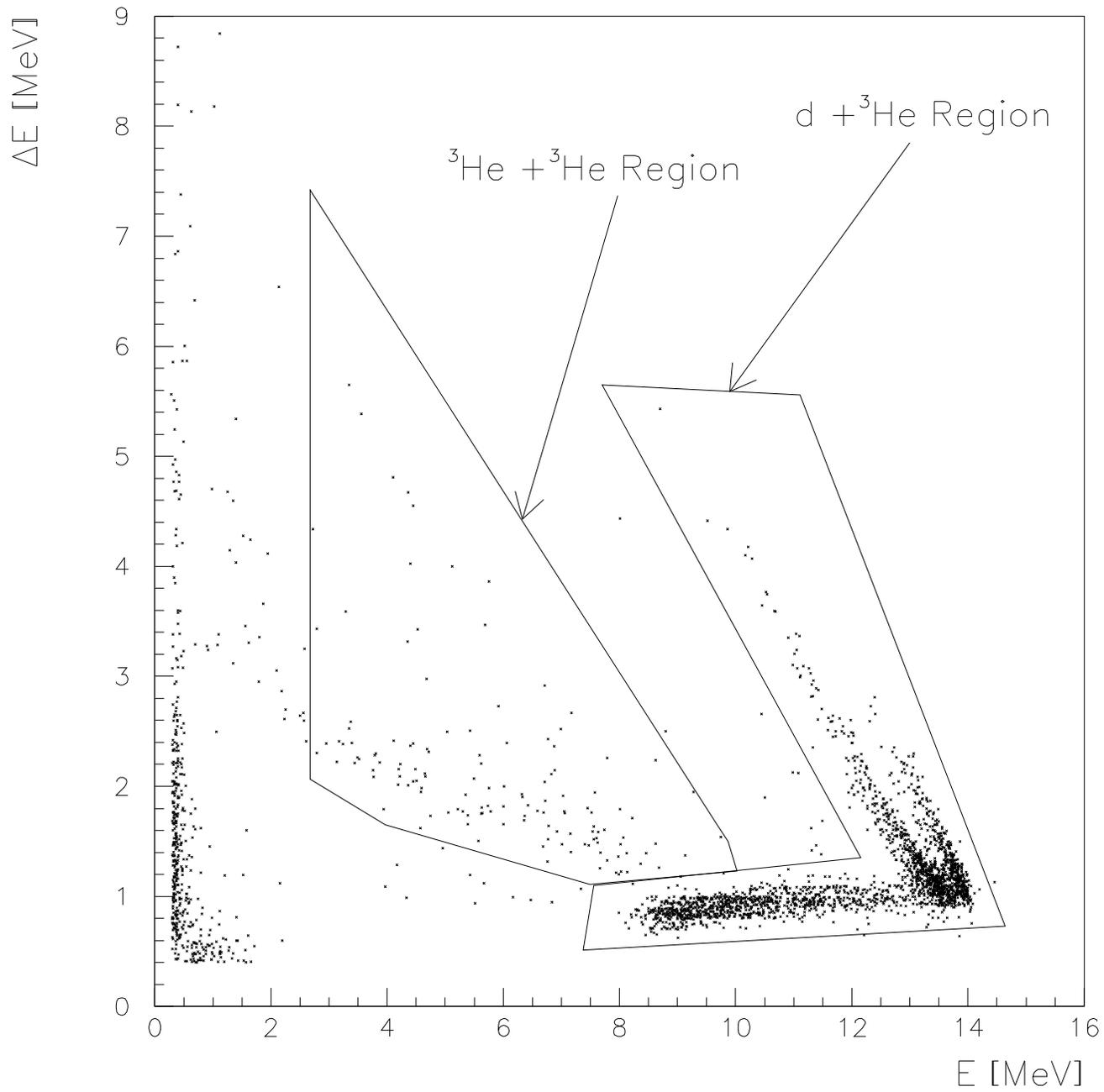
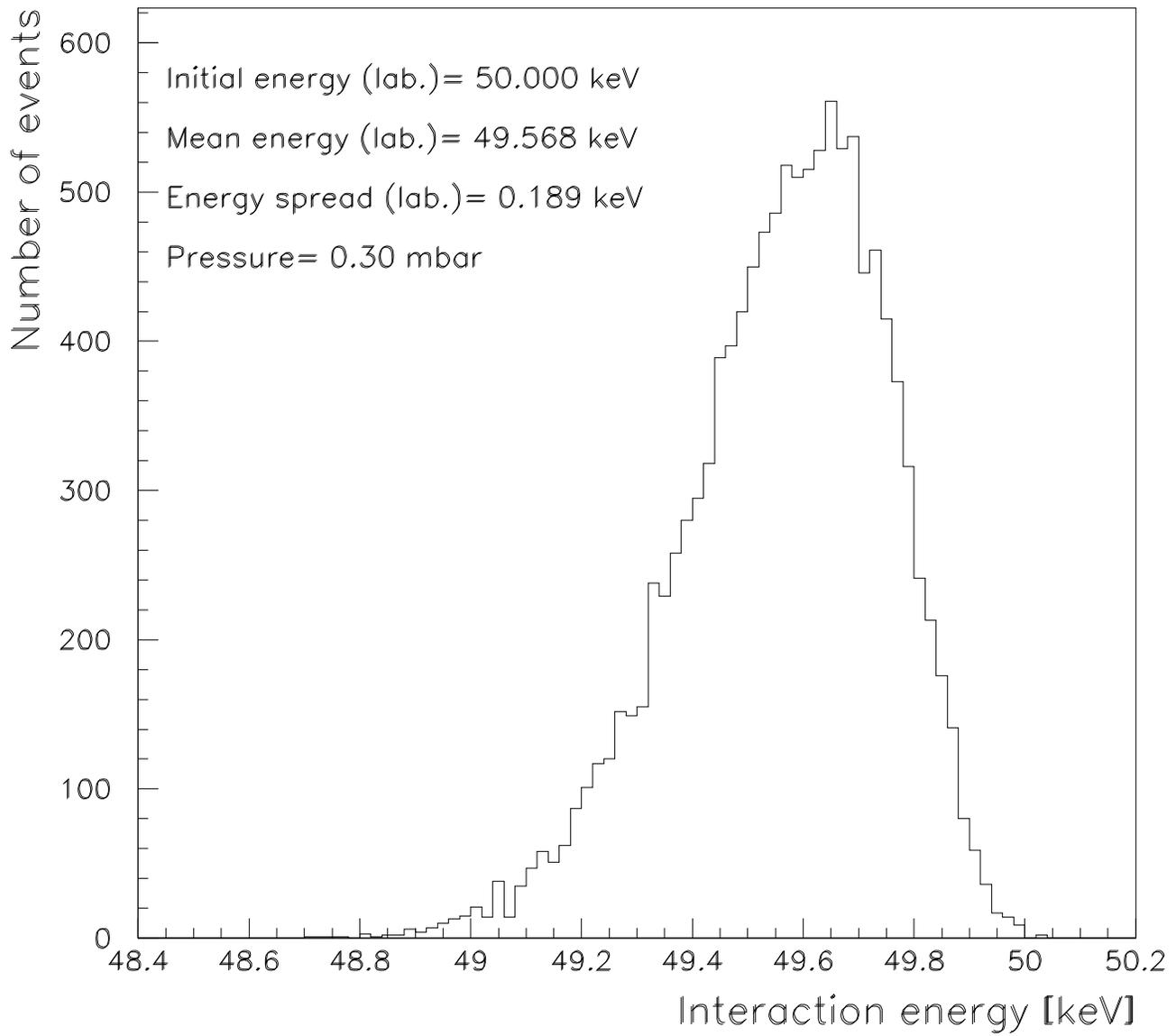
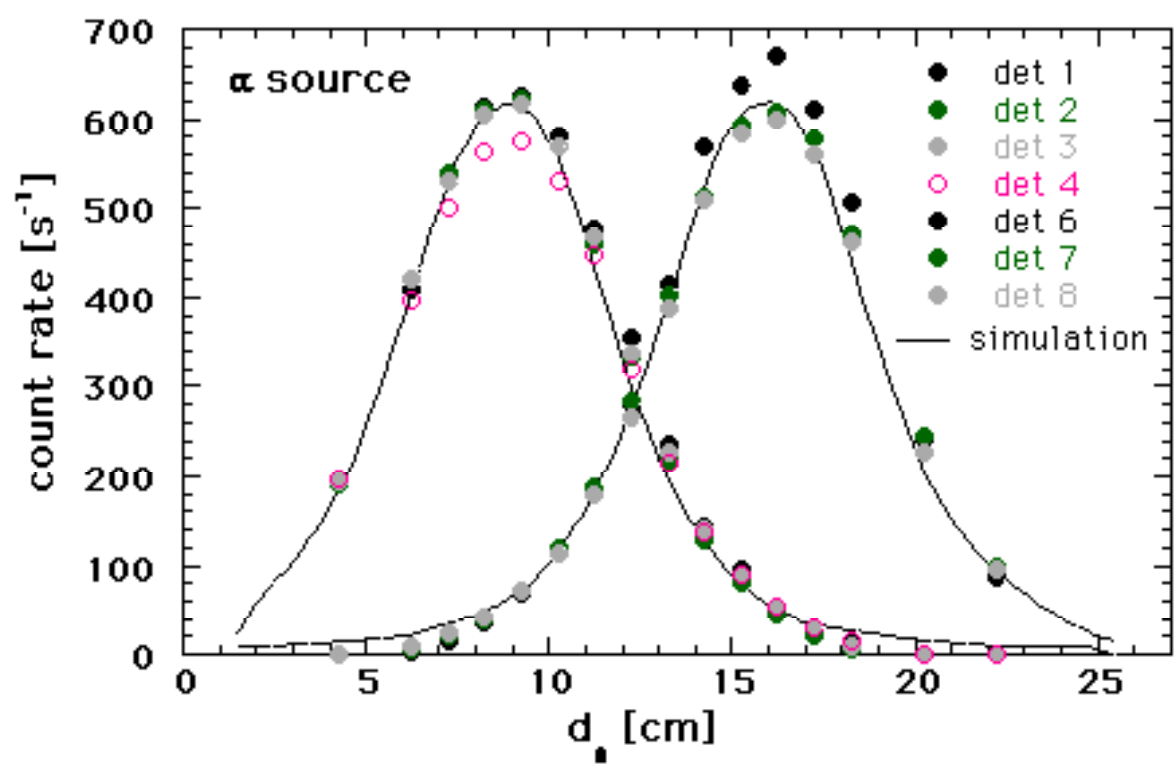
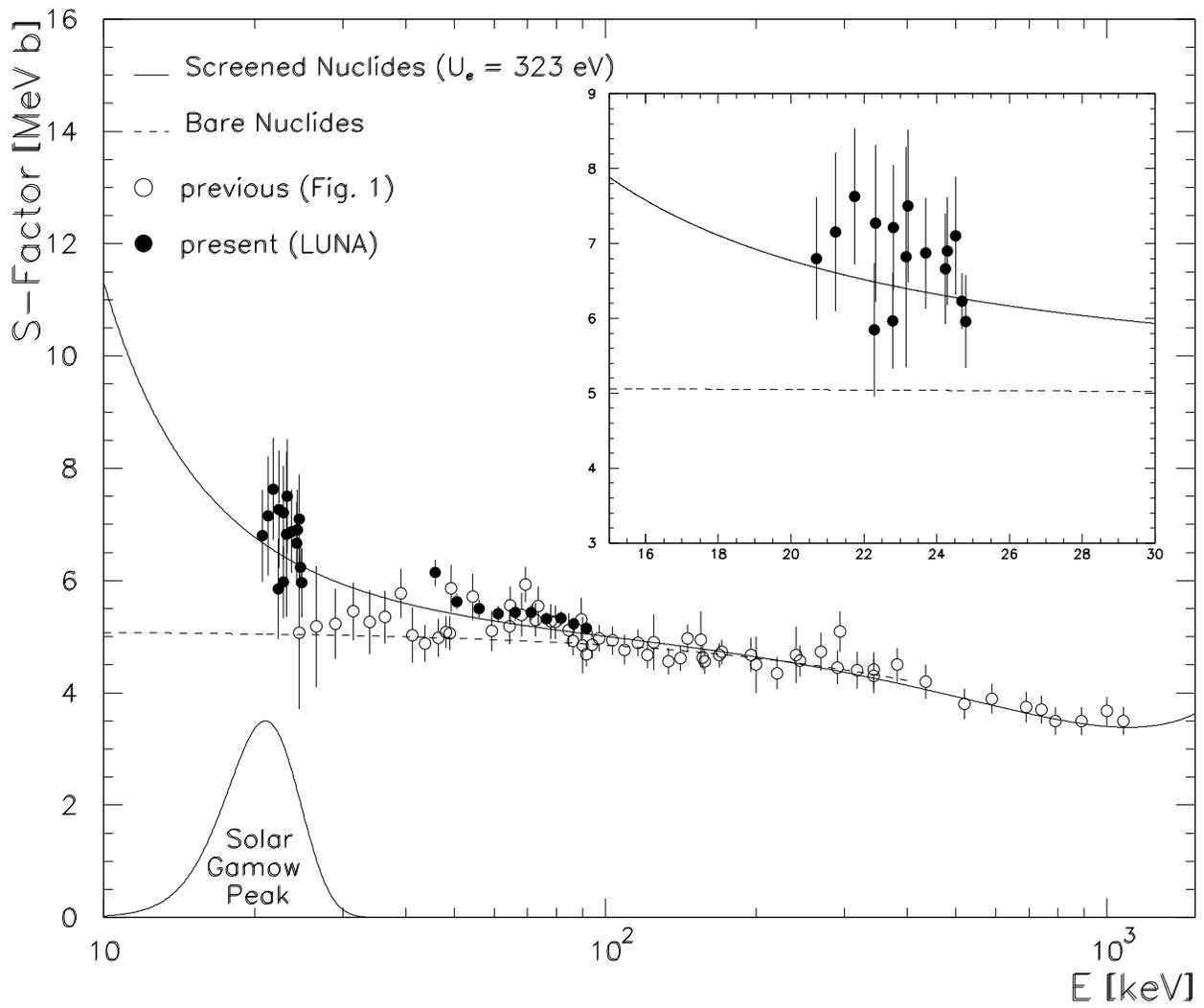
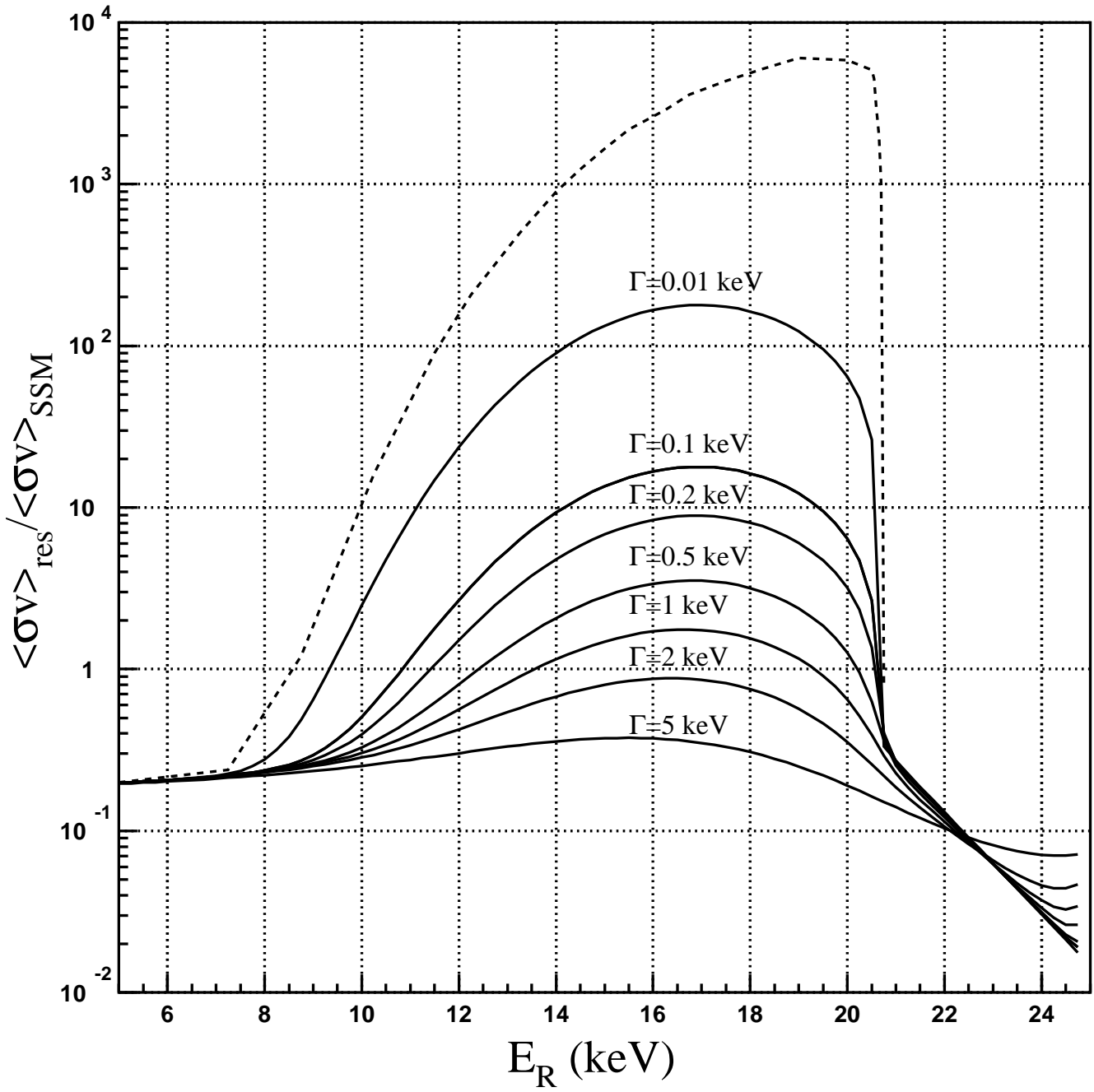
\end{document}